\documentclass{article}
\usepackage{iclr2025_conference,times}

\usepackage{amsmath,amsfonts,bm}

\def\eqref#1{equation~\ref{#1}}
\def\1{\bm{1}}

\DeclareMathAlphabet{\mathsfit}{\encodingdefault}{\sfdefault}{m}{sl}
\SetMathAlphabet{\mathsfit}{bold}{\encodingdefault}{\sfdefault}{bx}{n}

\usepackage[hyperfootnotes=false]{hyperref}
\usepackage{url}
\usepackage{booktabs}
\usepackage{tabularx}
\newcolumntype{R}{>{\raggedleft\arraybackslash}X}
\usepackage{xcolor}
\usepackage{amsmath}
\usepackage{amssymb}
\usepackage{xspace}
\usepackage{graphicx}
\graphicspath{{fig/}}
\usepackage{tikz}
\usetikzlibrary{arrows.meta,positioning,fit,backgrounds,calc,shapes.geometric,shapes.symbols,plotmarks}
\usepackage{pgfplots}
\pgfplotsset{compat=1.16}
\usepgfplotslibrary{groupplots}
\usepackage{sansmath}
\usepackage{listings}
\usepackage[most]{tcolorbox}
\usepackage{pifont}

\makeatletter

\renewcommand{\paragraph}{\@startsection{paragraph}{4}{\z@}%
  {-0.8ex \@plus -0.2ex \@minus -0.1ex}%
  {-0.6em}%
  {\normalfont\normalsize\bfseries}}
\makeatother

\newcommand{\pa}[1]{#1}

\newcommand{\HarnessBin}{26.9}\newcommand{\HarnessPart}{61.6}
\newcommand{\BaseBin}{20.6}\newcommand{\BasePart}{54.8}
\newcommand{\DeltaBin}{+6.3}\newcommand{\DeltaPart}{+6.8}
\newcommand{\Rgui}{1.1}
\newcommand{\RguiTurns}{11}
\newcommand{\GuiTasks}{28}
\newcommand{\SubPerTask}{4.2}%
\newcommand{\CostTask}{7.8}
\newcommand{\OrchTurnsTask}{57}
\newcommand{\SubTurnsTask}{98}
\newcommand{\TotalTurnsTask}{155}%
\newcommand{\OutTokTask}{100}%
\newcommand{\BaseCost}{72}
\newcommand{\BaseOutTok}{224}%
\newcommand{\CostRatio}{9}
\newcommand{\BinarySuccessN}{29}
\newcommand{\NonPerfect}{79}
\newcommand{\GateNonPerf}{76}
\newcommand{\GateNoFin}{3}
\newcommand{\GatePass}{68}
\newcommand{\GateFail}{8}
\newcommand{\GateCorrectPass}{28}
\newcommand{\GateCorrectFail}{1}
\newcommand{\Ceiling}{20}
\newcommand{\Gptbin}{13.0}\newcommand{\Gptpart}{49.5}
\newcommand{\Opseven}{18.2}\newcommand{\Opsevenp}{48.9}
\newcommand{\Snmaxbin}{8.3}\newcommand{\Snmaxpart}{41.5}
\newcommand{\Glmbin}{0.0}\newcommand{\Glmpart}{9.0}
\newcommand{\Mmbin}{4.6}\newcommand{\Mmpart}{22.3}
\newcommand{\Qwbin}{2.8}\newcommand{\Qwpart}{21.5}
\newcommand{\GptCost}{25.5}\newcommand{\OpsevenCost}{33.6}
\newcommand{\Bpart}{57.6}
\newcommand{\Cpart}{57.9}
\newcommand{\NoActBin}{18.5}\newcommand{\NoActPart}{51.3}
\newcommand{\NoVerBin}{23.1}\newcommand{\NoVerPart}{57.5}
\newcommand{\NoSusBin}{21.3}\newcommand{\NoSusPart}{58.7}
\newcommand{\CodeOnlyBin}{15.7}\newcommand{\CodeOnlyPart}{45.9}
\newcommand{\TransBbBin}{11.1}\newcommand{\TransBbPart}{42.0}
\newcommand{\TransBbBaseBin}{8.3}\newcommand{\TransBbBasePart}{41.5}
\newcommand{\TransBmBin}{78.4}\newcommand{\TransBmPart}{81.9}
\newcommand{\TransBmBaseBin}{77.3}\newcommand{\TransBmBasePart}{80.9}
\newcommand{\FaReason}{38}\newcommand{\FaVerif}{14}
\newcommand{\FaRecov}{52}\newcommand{\FaHard}{22}\newcommand{\FaOther}{5}

\newcommand{\ours}{StateAct\xspace}
\newcommand{\gate}{finish gate\xspace}

\definecolor{lstkw}{RGB}{0,90,180}
\definecolor{lststr}{RGB}{170,60,30}
\definecolor{lstcom}{RGB}{110,120,130}
\definecolor{lstnum}{RGB}{120,60,160}
\lstdefinestyle{cua}{basicstyle=\ttfamily\scriptsize,breaklines=true,
  keywordstyle=\color{lstkw}\bfseries,commentstyle=\itshape\color{lstcom},
  stringstyle=\color{lststr},numberstyle=\color{lstnum},
  numbers=none,frame=none,columns=fullflexible,showstringspaces=false,
  emphstyle=\color{lstkw}\bfseries}
\newcommand{\brand}[1]{\raisebox{-0.28ex}{\includegraphics[height=1.7ex]{#1}}\,}

\title{\ours: Program State, before Pixels, for Long-Horizon Computer-Use Agents}

\author{
\textnormal{Yan Yang}$^{\dagger}$ \quad \textnormal{Xiangru Jian}$^{\dagger}$ \quad \textnormal{Ziyang Luo}$^{\dagger}$ \quad \textnormal{Zirui Zhao} \quad \textnormal{Yutong Dai} \\
\textnormal{Ziji Shi} \quad \textnormal{Hanshu Yan} \quad \textnormal{Jun Hao Liew} \quad \textnormal{Silvio Savarese} \quad \textnormal{Junnan Li} \\[4pt]
\textbf{Salesforce AI Research} \\
}

\iclrfinalcopy

\makeatletter
\def\@maketitle{\vbox{\hsize\textwidth
{\LARGE\sc \@title\par}
\vskip 0.2in
\ificlrfinal\lhead{}\else\lhead{Under review as a conference paper at ICLR 2025}\fi
\def\And{\end{tabular}\hfil\linebreak[0]\hfil
        \begin{tabular}[t]{c}\bf\rule{\z@}{24pt}\ignorespaces}%
\def\AND{\end{tabular}\hfil\linebreak[4]\hfil
        \begin{tabular}[t]{c}\bf\rule{\z@}{24pt}\ignorespaces}%
\centering
\begin{tabular}[t]{c}\bf\rule{\z@}{24pt}\@author\end{tabular}%
\vskip 0.3in minus 0.1in}}
\makeatother

\begin{document}
\maketitle

\renewcommand{\thefootnote}{\fnsymbol{footnote}}
\footnotetext[2]{Main contributors.}
\renewcommand{\thefootnote}{\arabic{footnote}}

\begin{abstract}
Computer-use agents are usually improved by strengthening perception: better models for reading a screenshot and choosing where to click. Yet a screenshot is only a lossy rendering of the underlying \emph{program state}, \textit{e.g.}, the files, application backends, and DOM that hold the task data. Different states can produce the same pixels, while code can inspect and modify that state directly. \ours is a code-first, multi-agent harness built around this distinction. Its main agent works directly with program state by using code, while a dedicated GUI subagent handles screenshot-and-click interaction on the few subgoals that need it, just $\GuiTasks$ of $108$ tasks and $\Rgui\%$ of main-agent steps. The same direct access to program state also supports verification: an independent \gate double-checks the saved result for structural failures, \textit{e.g.,} output that is missing, unsaved, or written to the wrong path. To stay on track over hundreds of steps, the main agent hands subgoals to fresh subagents, keeping its own context focused. On OSWorld~2.0, \ours lifts Claude~Opus~4.8 from $\BaseBin\%$ to $\HarnessBin\%$ on binary success, and from $\BasePart\%$ to $\HarnessPart\%$ on partial success, at $\sim\CostRatio\times$ lower cost per task than the same model driven by screenshots alone; a code-only variant with no GUI subagent reaches only $\CodeOnlyPart\%$ partial, below that screenshot-based baseline's $\BasePart\%$. In general, grounding action, verification, and memory in state, what we call state-grounding, shifts the main bottleneck from perception toward reasoning: failures depend more on what the agent \emph{thinks} than on what it \emph{sees}.
\end{abstract}

\section{Introduction}
\label{sec:intro}

Computer-use agents operate the real desktop software people use every day, \textit{e.g.}, file managers, spreadsheets, calendars, browsers, email. Most current work improves these agents by strengthening screen perception: better models for reading and acting on rendered pixels~\citep{uitars,aguvis,gta1,setofmark}. On \emph{long-horizon} tasks, however, reading the screen is only part of the problem. Agents must also carry out many dependent operations and determine whether the final result is correct and saved. \ours therefore makes \emph{program state} (\textit{i.e.}, the files, application backends, and DOM that hold the task data) as the main interface for agents, while retaining visual interaction through a dedicated GUI specialist. We call this \emph{state-grounding}.

The key idea is simple: a desktop task is judged by the state it leaves behind, which a screenshot alone may not reveal. For example, a spreadsheet can display the same total whether a cell contains a formula or a literal, and relevant rows may be hidden or off-screen. Pixels alone hide these task-critical distinctions; direct state access exposes them and edits the values behind them.
This gap widens with task length. A single lossy read is usually harmless, but a per-step proxy compounds: over hundreds of dependent operations, small misreads accumulate into a wrong deliverable.
Worse, a screenshot offers no dependable signal that the final artifact is correctly complete, exactly the judgment a long-horizon task hinges on.
\ours therefore acts on program state, checks structural completion against the persisted artifact, and uses context management to carry task facts and plans across the long horizon.

\begin{figure}[t]
\centering
\includegraphics[width=\linewidth]{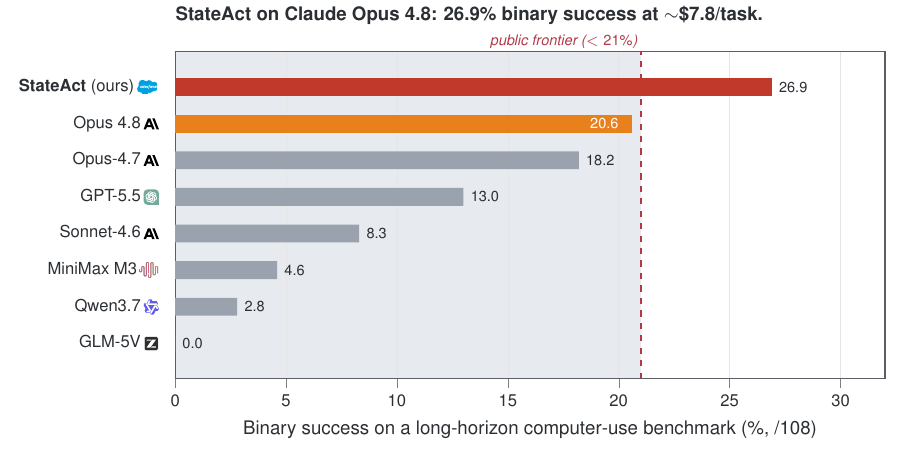}
\vspace{-1.5em}
\caption{$\HarnessBin\%$ binary success on Claude~Opus~4.8: the only entry above the $<21\%$ public frontier, at $\sim\CostRatio\times$ lower cost than the same backbone's reference computer-use-agent harness ($\sim\$\CostTask$ \textit{vs.}\ $\sim\$\BaseCost$ per task). Orange marks the Claude~Opus~4.8 reference baseline.}
\label{fig:headline}
\end{figure}

\begin{figure}[t]
\centering
\includegraphics[width=.9\linewidth]{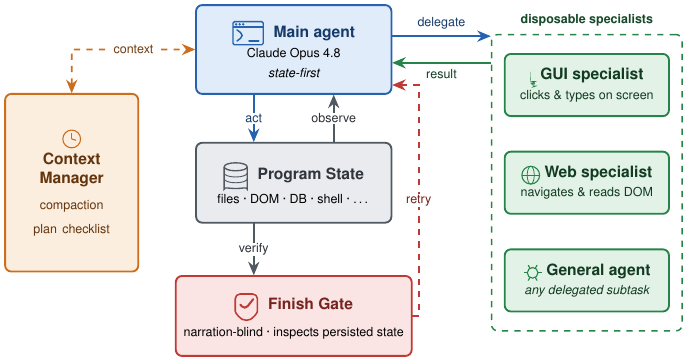}
\vspace{-1em}
\caption{\ours architecture. The main agent works directly with program state and uses fresh-context GUI and browser specialists for complementary interaction. The \gate independently checks persisted state, while the context manager preserves the plan and task facts over hundreds of steps. The gate fires on \texttt{finish} (after at least three non-finish steps), for up to three rounds.}
\label{fig:arch}
\end{figure}

\ours (Figure~\ref{fig:arch}) builds the full agent loop around three parts: state-grounded action, verification, and long-horizon execution.
For action, the main agent reads and writes the task's real artifacts through persistent \texttt{bash}, Python, and a file editor. Code alone is not enough, since not every application exposes a comprehensive API or otherwise accessible state. Empirically, with only \texttt{bash} the same agent scores just $\CodeOnlyPart\%$ partial success, below the $\BasePart\%$ vision baseline, so a dedicated GUI subagent supplies screen-based interaction, used on just $\GuiTasks$ of $108$ tasks and $\Rgui\%$ of main-agent steps. For verification, a separate gate checks the persisted result without seeing the main agent's account of its work.
For long-horizon execution, a growing context of a single agent fills with stale detail and loses task facts over hundreds of steps, so applying fresh-context delegation, compaction, and an externalized plan keeps the main agent focused.

We make three contributions.
i)~\emph{Architecture}: \ours makes the main agent observe and act primarily on program
state, retains a dedicated GUI subagent for visual interaction
($\Rgui\%$ of main agent steps, $\GuiTasks$ of $108$ tasks), and carries the same
discipline into verification and long-horizon memory
(\S\ref{sec:principle}, \S\ref{sec:method}).
ii)~\emph{Empirical gain}: on the strongest backbone (Claude~Opus~4.8), the
harness alone lifts binary success $\BaseBin\%\!\to\!\HarnessBin\%$ and partial
$\BasePart\%\!\to\!\HarnessPart\%$ over the reference, exceeding every public entry on a
system-level comparison (Figure~\ref{fig:headline} and Table~\ref{tab:taxonomy}).
iii)~\emph{Diagnosis}: we identify what drives the gain and what limits it.
The gain comes from \emph{what} the agent observes (state), not from added agentic depth
(offered recursion fires on only $7$ of $108$ tasks and never nests;
\S\ref{sec:exp-ablation}). The limit is value correctness: a state-grounded verifier
without a ground-truth oracle catches structural errors but still passes $\GatePass$ of
the $\GateNonPerf$ non-perfect tasks that reached the gate, on value errors it cannot
re-derive independently, \textit{i.e.}, dominantly reasoning errors (\S\ref{sec:diagnostic}).

\section{Related Work}
\label{sec:related}

\paragraph{Grounding-based computer-use agents.}
A large body of work improves the perceptual channel: native GUI action models (\pa{UI-TARS~\citep{uitars}}, \pa{Aguvis~\citep{aguvis}}), test-time selection (\pa{GTA1~\citep{gta1}}), and easier-to-ground observations such as \pa{Set-of-Mark~\citep{setofmark}}. These make the agent better at reading the screen; \ours is complementary but inverts the default: it grounds in
program state and treats direct screen interaction as a delegated fallback.

\vspace{-.5em}
\paragraph{Code and hybrid action spaces.}
Acting through code rather than a UI-shaped schema is well established: \pa{CodeAct~\citep{codeact}} makes executable code the action space, and OSWorld's native action space is Python via \texttt{pyautogui}. \pa{CoAct-1~\citep{coact1}} pairs a programmer with a co-equal GUI operator under an orchestrator; \pa{UltraCUA~\citep{ultracua}} and \mbox{\pa{ComputerRL~\citep{computerrl}}} learn hybrid API/GUI action spaces via reinforcement learning; \pa{UFO2~\citep{ufo2}} fuses a unified GUI--API layer; and \pa{SWE-agent~\citep{sweagent}} argues agents deserve purpose-built interfaces. The architectural distinction is one of default routing: \ours keeps code/state operations in the main agent and delegates direct GUI control to a subagent, while extending the same state inspection to its verification gate (\S\ref{sec:gate}).

\vspace{-.5em}
\paragraph{Self-verification for agents.}
Grounding verification in system state rather than model narration is likewise established. \pa{OpenComputer~\citep{opencomputer2026}} uses \emph{hard-coded}, per-application state verifiers that align with human adjudication more closely than an LLM judge; \pa{Agentic Reward Modeling~\citep{vagen2026}} probes hidden system state via proactive interaction; \pa{MCPWorld~\citep{mcpworld}} verifies via backend instrumentation. Our \gate is deliberately \emph{weaker on value but stronger on generality}: it is a single prompt-level check that applies to all $108$ tasks with \emph{zero} per-task verifier engineering, trading OpenComputer's oracle-grade value checking for a structural-only ceiling we measure and own (\S\ref{sec:diagnostic}). We also distinguish the \gate from iterative self-correction. \pa{Reflexion~\citep{reflexion}} and \pa{Self-Refine~\citep{selfrefine}} have the agent reflect on its \emph{own} narrated trajectory, making self-consistency their substrate. Our \gate is the opposite: a fresh, independent, third-party check at a single terminal boundary that \emph{refuses} to read the agent's narration and directly inspects the persisted deliverable. This is closer to an actor--critic separation or an independent acceptance test than to reflection. This matters because narration-conditioned judges have been shown to be biased by the behavior they observe~\citep{agreementbias}. Designing narration out is the point, and \S\ref{sec:diagnostic} measures how far it gets us.

\section{The State-Grounding Principle}
\label{sec:principle}

\paragraph{A model of the two channels.}
Let $s\in\mathcal{S}$ be the computer program state (\textit{e.g.}, files, application backends, DOM, tables). An agent never observes $s$ directly; it observes
it through a channel. The \emph{pixel channel} is the render map
$o_{\mathrm{pix}}=f_{\mathrm{render}}(s)$, and the \emph{state channel} is a query
map $o_{\mathrm{state}}=g(s)$ (\textit{e.g.}, a shell command, a workbook read, a DOM
serialization). Two structural facts drive \ours.

\begin{figure}[t]
\centering
\includegraphics[width=\linewidth]{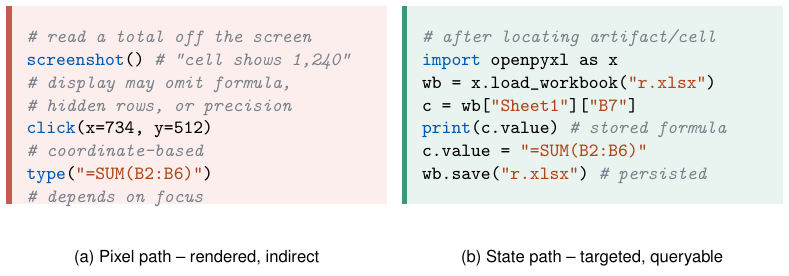}
\vspace{-2em}
\caption{The same subgoal via each channel. The pixel channel yields a rendered, ambiguous value at a fragile coordinate; the state channel reads and writes the exact artifact the task's deliverable is made of. \ours uses (b) for state-addressable operations and a dedicated GUI subagent for (a) when visual interaction is required.}
\label{fig:statevspixel}
\end{figure}

\emph{i) Rendering is lossy and non-injective.} Many distinct states render to indistinguishable, or arbitrarily close, screenshots (Figure~\ref{fig:statevspixel}, a displayed total may hold a literal or a formula, be rounded, or be scrolled off-screen). Formally $f_{\mathrm{render}}$ is not injective, so $f_{\mathrm{render}}^{-1}$ does not exist and no perceptual model, however good, can in general recover $s$ from $o_{\mathrm{pix}}$. Over the substate a task touches, the state channel $g$ is effectively invertible. For example, in Figure~\ref{fig:statevspixel}~(b), from \texttt{load\_workbook(...)["B7"].value} the agent recovers the formula exactly.
The loss is benign in one shot but compounds over a \emph{long horizon}: a per-step lossy
proxy accumulates state drift across hundreds of steps, and the state channel's advantage grows with task length, which is the regime this paper targets.

\emph{ii) The deliverable \emph{is} the state.} A desktop task deliverable is a change to program state, not a screenshot of it. Whether a task succeeded is determined by the program state (\textit{e.g.}, whether the user goal is met in the saved file or application settings). Formally, let $G(s)$ denote the success predicate over state. For tasks whose deliverable depends on non-rendered content (formulas, hidden rows, off-screen data, backend state), $G(s)$ cannot be recovered from $f_{\mathrm{render}}(s)$ alone (there is no $\tilde G$ with $G(s)=\tilde G(f_{\mathrm{render}}(s))$ for all such $s$), because rendering discards that content. A GUI may still reach such content when the interface exposes it, but usually only through multiple steps, such as selecting a cell to reveal its formula or scrolling hidden rows into view; the state channel reads and writes it directly and exactly. For tasks that only require reading and summarizing information (no state change), either channel can work; the distinction matters only when the deliverable depends on content the screen does not show. For such tasks, acting and checking on $g(s)$ operates on the \emph{actual artifact the task asks for}, whereas acting on $f_{\mathrm{render}}(s)$ operates on a lossy image of it.

\vspace{-.5em}
\paragraph{Boundary: where the state channel does not help.}
The model delimits the principle's scope. When a task targets a visual outcome
(\textit{e.g.}, image editing, layout, chart appearance, WYSIWYG output), the rendered screen is the relevant observation, and the state channel gives no leverage. More generally, any subgoal that can only be expressed as a
rendered interaction (dragging a canvas element, dismissing a non-scriptable modal, reading
a value that exists only on-screen) falls outside the state channel's reach. We
accordingly distinguish three conceptual task types: \emph{state-addressable} (\textit{i.e.}, a code path
to the target artifact exists), \emph{hybrid} (\textit{i.e.}, mostly state, one irreducibly visual
subgoal), and \emph{render-only} (\textit{i.e.}, pixel/appearance judgments). The principle predicts
state-grounding helps the first two and gives no advantage on the third.
In our evaluation (\S\ref{sec:experiments}), the per-capability breakdown
(Table~\ref{tab:taxonomy}) is qualitatively consistent: \ours's margins are largest on state-addressable capabilities and smallest on the most render-only ones, human-in-the-loop and multimodal editing.

\section{\ours}
\label{sec:method}

\begin{figure}[t]
\centering
\includegraphics[width=\linewidth]{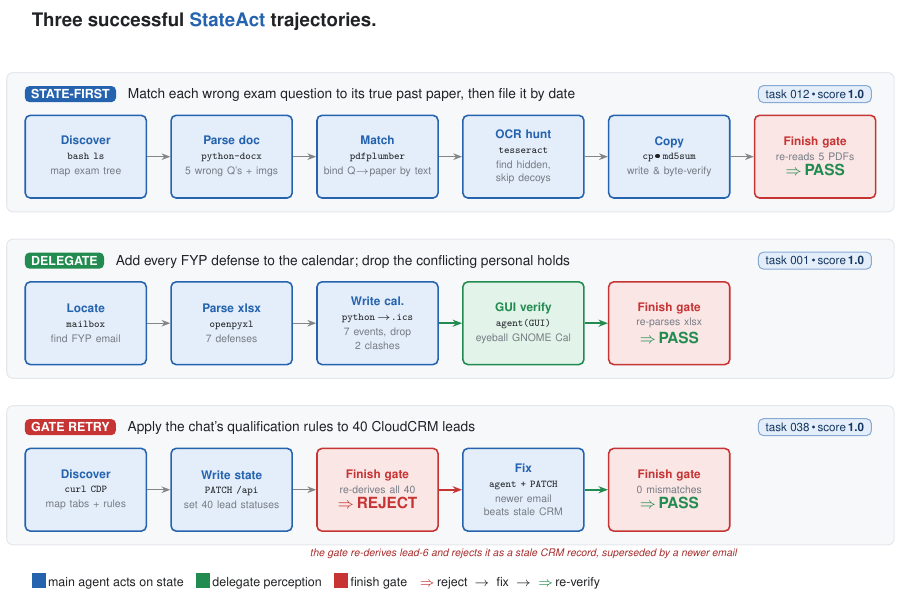}
\vspace{-1em}
\caption{Three successful trajectories selected to illustrate state-first action,
GUI delegation, and a gate retry (chips = key steps, color-coded by role; all
reach score $1.0$).
{State-first} (task~012): the main agent acts entirely in
code, using \texttt{python-docx}/\texttt{pdfplumber}/\texttt{tesseract} to bind each
wrong exam question to its true past paper, correctly rejecting the distractor questions, with no
GUI at all. {Delegate} (task~001): code writes the calendar \texttt{.ics}
directly and a GUI specialist is invoked \emph{only} to eyeball GNOME Calendar.
{Gate retry} (task~038): the \gate re-derives all 40 CRM lead
statuses from source and rejects lead-6 as sourced from a stale record; the agent re-derives it
from a newer email that supersedes the stale record, and re-verification passes.}
\label{fig:traj}
\end{figure}

\ours (Figure~\ref{fig:arch}) has three components:
i)~a main agent that acts through code on program state,
ii)~an independent \gate that verifies completion by re-reading the real artifacts,
and iii)~context management that sustains the run over hundreds of steps.
Figure~\ref{fig:traj} shows example trajectories.

\begin{table}[t]
\caption{Per-capability performance on OSWorld~2.0's ten capability labels, sorted by \ours; each cell is {binary\,/\,mean-partial} (\%). \ours (bold) is our state-grounding harness on Claude~Opus~4.8; \emph{Opus~4.8 (ref.)} is the same backbone under the computer-use-agent (CUA) harness. For Opus-4.7, we report only its best reported batched configuration. ``\,--\,'' denotes not available.}
\label{tab:taxonomy}
\vspace{-.5em}
\begin{center}\footnotesize\setlength{\tabcolsep}{2.1pt}
\resizebox{\textwidth}{!}{%
\begin{tabular}{lcccccccc}
\toprule
& \multicolumn{8}{c}{\textbf{Binary\,/\,mean-partial score (\%)}} \\
\cmidrule(lr){2-9}
{Capability} & \brand{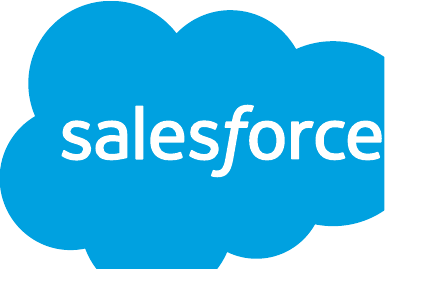}\textbf{\ours} & Opus~4.8 & \brand{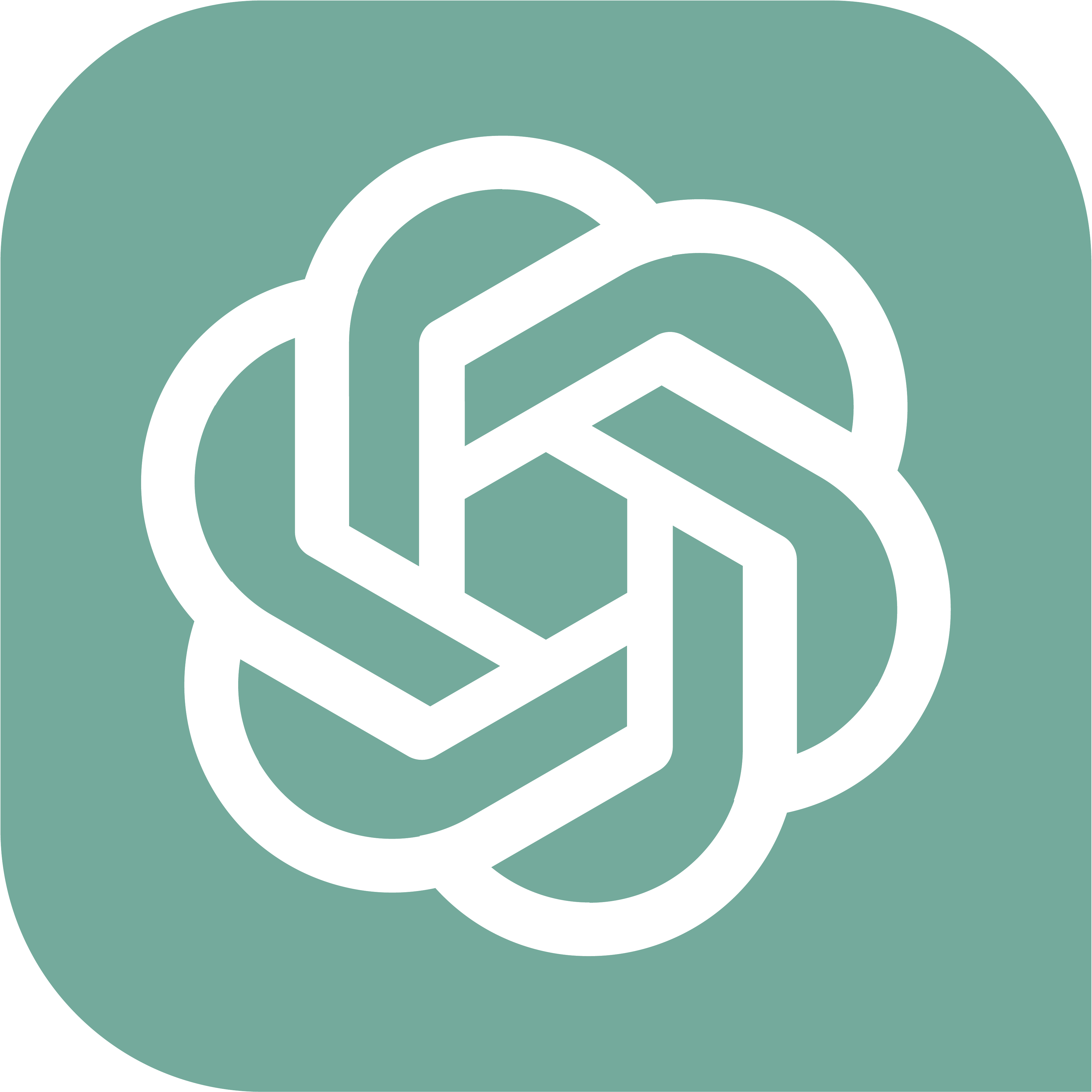}GPT-5.5 & \brand{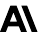}Opus-4.7 & \brand{anthropic}Sonnet-4.6 & \brand{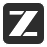}GLM-5V & \brand{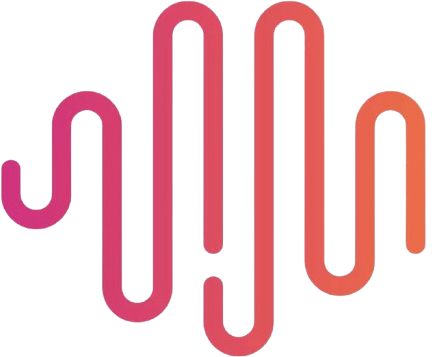}MiniMax M3 & \brand{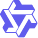}Qwen3.7 \\
 & {\scriptsize(Opus~4.8)} & {\scriptsize(ref.\ CUA)} & & & & & & \\
\midrule
Multi-item state   & \textbf{27.9/66.7} & -- & 14.0/50.6 & -- & 11.6/46.7 & 0.0/11.2 & 7.0/23.2 & 2.3/20.2 \\
Streaming          & \textbf{66.7/66.7} & -- & 50.0/57.8 & -- & 0.0/4.7 & 0.0/1.2 & 0.0/6.4 & 0.0/0.0 \\
Cross-source       & \textbf{26.1/64.9} & -- & 13.0/52.4 & -- & 10.9/45.8 & 0.0/11.7 & 6.5/24.9 & 6.5/26.3 \\
Conflict disambig. & \textbf{30.8/64.5} & -- & 20.5/51.4 & -- & 12.8/42.4 & 0.0/12.4 & 7.7/24.3 & 7.7/29.1 \\
Visual-spatial     & \textbf{31.1/61.3} & -- & 11.1/51.2 & -- & 8.9/36.5 & 0.0/4.6 & 4.4/19.8 & 2.2/19.8 \\
Implicit state     & \textbf{30.2/60.1} & -- & 14.0/47.3 & -- & 9.3/37.0 & 0.0/11.0 & 4.7/24.4 & 2.3/24.1 \\
Tutorial           & \textbf{13.6/55.3} & -- & 9.1/37.5 & -- & 13.6/43.5 & 0.0/7.8 & 4.5/15.0 & 4.5/15.7 \\
Multimodal edit    & \textbf{20.0/54.2} & -- & 6.7/47.0 & -- & 6.7/37.5 & 0.0/6.2 & 6.7/22.3 & 0.0/20.6 \\
Dynamic env.       & \textbf{30.0/53.5} & -- & 30.0/46.2 & -- & 10.0/22.0 & 0.0/5.8 & 0.0/17.9 & 0.0/16.3 \\
Human-in-the-loop & \textbf{0.0/43.9} & -- & 16.7/43.1 & -- & 16.7/51.9 & 0.0/12.5 & 0.0/16.8 & 0.0/22.5 \\
\midrule
\emph{Overall (/108)} & \textbf{\HarnessBin/\HarnessPart} & \BaseBin/\BasePart & \Gptbin/\Gptpart & \Opseven/\Opsevenp & \Snmaxbin/\Snmaxpart & \Glmbin/\Glmpart & \Mmbin/\Mmpart & \Qwbin/\Qwpart \\
\bottomrule
\end{tabular}}
\end{center}
\end{table}

\subsection{Act on state}
\label{sec:pillar-act}
The main agent action space is code and structured operations: persistent
\texttt{bash}, a file editor, a read-only \texttt{view\_image} for
image \emph{files}, a \texttt{plan} checklist, a \texttt{finish} action, and an
\texttt{agent} delegation tool. No live screen actuation (\textit{e.g.}, mouse or keyboard) is exposed to the main agent.

\vspace{-.5em}
\paragraph{State discovery.}
The main agent acts to find where an application persists its state, drawing on two signals: the model own priors about how common desktop applications store state on disk
(\textit{e.g.}, mail stores, office-document formats, browser profiles, application databases), and active
probing (\textit{e.g.}, \texttt{find}/\texttt{ls}/\texttt{grep}/\texttt{sqlite3}) when the prior is
uncertain. Discovery locates \emph{where} state lives, never \emph{what} the target value is.

\vspace{-.5em}
\paragraph{Delegation rule.}
The main agent uses the dedicated \texttt{cua} (GUI) subagent when a subgoal is
\emph{irreducibly visual}: i)~no file/backend/DOM path to the target could be found by
probing, or ii)~the effect is only expressible as a rendered interaction (dragging a canvas selection, dismissing a non-scriptable modal, or reading a value that exists only on-screen).
The measured GUI-use rate is $\Rgui\%$ of main agent steps ($\GuiTasks$ of $108$ tasks touch the GUI subagent at least once), rising to $\sim\RguiTurns\%$ of total model turns once the subagent's interior turns are included.Browser work is handled by a dedicated \texttt{web} subagent that navigates, executes JavaScript, serializes the DOM to markdown, and clicks by CSS selector, grounding it in structured state rather than a screenshot.

\subsection{Verify on state}
\label{sec:gate}
\begin{figure}[t]
\centering
\vspace{-2em}
\includegraphics[width=\linewidth]{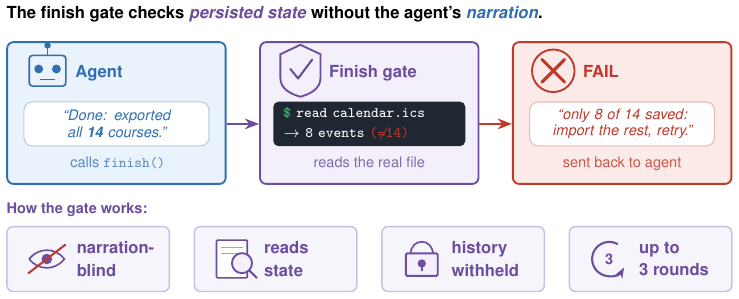}
\caption{An independent \gate transcript (condensed from the calendar-import failure class
in \S\ref{sec:diagnostic}). The gate checks the actual persisted store, catching that only
$8$ of $14$ events were saved (the \texttt{Downloads} copy was complete but the store was
not). This is the \emph{structural} defect class the gate catches reliably;
\S\ref{sec:diagnostic} reports the class it does not.}
\label{fig:gate}
\end{figure}

When the main agent calls \texttt{finish} (permitted only after at least three non-finish steps), an independent \gate
(Figure~\ref{fig:gate}) spawns. Its context contains \emph{only} the verbatim task instruction and machine access (\texttt{bash}, file reads, DOM queries), with editor mutations blocked. It never sees the main agent's message history, plan, or finish rationale, nor the expected values. This gives the gate four properties.
i)~{Narration-blind}: it sees only the task and the machine, not the agent's
claims; unlike a Reflexion-style self-critique, it cannot be talked into agreement by the
trajectory's own story.
ii)~{State-grounded}: it checks the requested result in the exact persisted artifact
(\textit{e.g.}, the calendar store, the app backend, the saved spreadsheet's cell formulas),
rather than relying on the UI.
iii)~{Anti-capture}: it must independently locate the real deliverable the task
names (\textit{e.g.}, the exact file or backend the instruction asks to change) and ground its checks there; evidence found only in a side file the agent created itself, rather than in the deliverable the task names, is rejected.
iv)~{Bounded correction}: on rejection, the main agent is sent back to fix the
named gap, up to three rounds. We report the gate's measured operating behavior in
\S\ref{sec:diagnostic}.

\subsection{Sustain state}
\label{sec:pillar-sustain}
Long episodes (up to $200$ main agent turns, plus a mean of $\SubPerTask$ subagent delegations per task, each its own $\leq$50-turn loop) would overflow a naive context. Three mechanisms
address this.
i)~{Fresh-context specialists}: each delegation runs in its own context window, seeded with a focused subtask and returning a concise report, so image-heavy or exploratory subwork stays out of the main agent's state model.
ii)~{Auto-compaction}: near the context limit, the oldest prefix is summarized at an assistant boundary and images are stripped, preserving state facts while reclaiming context budget.
iii)~{Externalized plan}: a task checklist persisted outside the message history is
re-injected each turn, surviving compaction and anchoring multi-part tasks.

\section{Experiments}
\label{sec:experiments}

\subsection{Experimental setup}
\label{sec:exp-setup}
We evaluate \ours on OSWorld~2.0~\citep{osworld2}, a standard long-horizon GUI benchmark
(108 tasks). Following the benchmark's protocol, we report binary success,
mean partial score (fractional credit), and cost per task
(USD). All runs use Claude~Opus~4.8 with adaptive thinking and a $200$-turn main-agent budget per agent invocation.

\vspace{-.5em}
\paragraph{Turn accounting.} We distinguish model turns from tool uses. The main agent
averages $\sim\OrchTurnsTask$ model turns per episode; each subagent delegation adds
$\sim$23 interior turns (capped at 50), for $\sim\TotalTurnsTask$ model turns per
task ($\sim\OrchTurnsTask$ main agent $+$ $\sim\SubTurnsTask$ subagent). We use that
total for the cross-system comparison in Table~\ref{tab:efficiency}.

\subsection{Main result}
\label{sec:mainresult}
On the identical backbone (Table~\ref{tab:taxonomy}), replacing the computer-use-agent
harness with \ours raises binary success by $\DeltaBin$ points and mean partial by
$\DeltaPart$, while cutting output tokens ($\BaseOutTok\text{K}\to\OutTokTask\text{K}$)
and dollar cost (from $\sim\$\BaseCost$~\citep{osworld2} to $\sim\$\CostTask$ per task, a $\sim\CostRatio\times$ reduction).
\ours ($\HarnessBin\%$~/~$\HarnessPart\%$) is the best-performing entry, exceeding the same-backbone reference.

\begin{figure}[t]
\centering
\vspace{-2em}
\includegraphics[width=\linewidth]{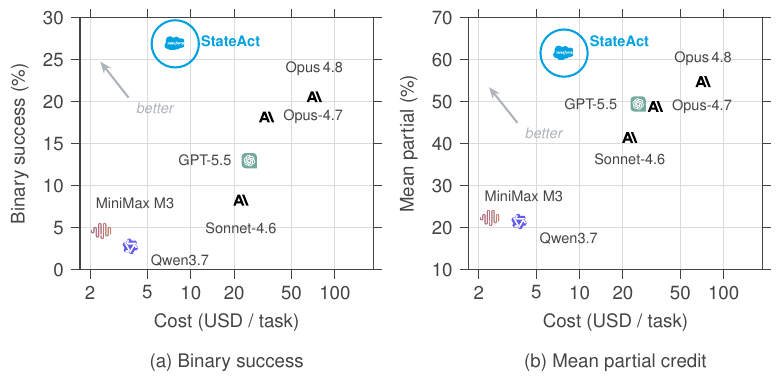}
\caption{Cost--accuracy frontier on OSWorld~2.0 (binary success \textit{vs.}\ USD per task).
Public points are from publicly released OSWorld~2.0 trajectories~\citep{osworld2}. \ours (red) sits in the top-left: more accurate and cheaper than every stronger public baseline shown; only the much weaker MiniMax M3 and Qwen entries cost less.}
\label{fig:pareto}
\end{figure}

On the cost--accuracy frontier (Figure~\ref{fig:pareto}), \ours sits alone in the top-left:
it exceeds the best public entry (Opus-4.7, $\Opseven\%$) by $8.7$ binary points and
GPT-5.5 ($\Gptbin\%$) by $13.9$, and adds $\ge$12 partial points over both, while costing $\sim\$\CostTask$ per task, below GPT-5.5 ($\sim\$\GptCost$) and Opus-4.7 ($\sim\$\OpsevenCost$)
and far below the same-model reference ($\sim\$\BaseCost$); among the displayed systems, only the much weaker MiniMax M3 and Qwen entries cost less.
Compared to the reference computer-use-agent harness on Claude~Opus~4.8, \ours is
both more accurate ($\DeltaBin$ / $\DeltaPart$ points higher) and $\sim\CostRatio\times$
cheaper, showing that state-grounding improves quality and cost simultaneously.

\begin{table}[t]
\caption{Mean model turns per task, by capability. For \ours: main agent plus subagent
interior turns ($\OrchTurnsTask{+}\SubTurnsTask{=}\TotalTurnsTask$). Public entries are 
from publicly released OSWorld~2.0 trajectories. ``\,--\,'' denotes per-task trajectory is not available. Capabilities are multi-label, so a task can count toward several columns; the per-capability means therefore do not average to \emph{Overall}, which is the per-task mean over all $108$ tasks.}
\label{tab:efficiency}
\begin{center}\footnotesize\setlength{\tabcolsep}{3pt}
\begin{tabular*}{\linewidth}{@{\extracolsep{\fill}}lccccccccccc@{}}
\toprule
{Model} & \rotatebox{50}{Stream} & \rotatebox{50}{Tutor} & \rotatebox{50}{Vis-sp} & \rotatebox{50}{M-edit} & \rotatebox{50}{Dyn} & \rotatebox{50}{Implic} & \rotatebox{50}{H-loop} & \rotatebox{50}{M-item} & \rotatebox{50}{X-src} & \rotatebox{50}{Confl} & \rotatebox{50}{Overall} \\
\midrule
\textbf{\ours}  & 170 & 173 & 137 & 136 & 253 & 180 & 282 & 163 & 193 & 184 & {155} \\
Opus~4.8         & -- & -- & -- & -- & -- & -- & -- & -- & -- & -- & 103 \\
GPT-5.5         & 152 & 98 & 85 & 78 & 111 & 105 & 72 & 107 & 102 & 93 & 95 \\
Opus-4.7        & -- & -- & -- & -- & -- & -- & -- & -- & -- & -- & 161 \\
Sonnet-4.6      & 418 & 319 & 229 & 237 & 288 & 284 & 198 & 267 & 270 & 235 & 253 \\
MiniMax M3       & 314 & 370 & 309 & 310 & 312 & 351 & 282 & 345 & 362 & 341 & 327 \\
GLM-5V          & 321 & 293 & 299 & 341 & 329 & 313 & 320 & 338 & 338 & 333 & 314 \\
\bottomrule
\end{tabular*}
\end{center}
\end{table}

\subsection{Ablation}
\label{sec:exp-ablation}

\paragraph{Component sensitivity (Table~\ref{tab:ablation}a).}
We remove one component at a time from full \ours: $-$act re-introduces the computer tool
(GUI in the main agent, \texttt{cua} blocked); $-$verify disables the \gate; $-$sustain
removes compaction and the plan tool. Removing act-on-state produces the largest drop
(partial $\HarnessPart\%\to\NoActPart\%$, below even the reference's $\BasePart\%$),
consistent with code-first action on state being the largest single contributor. Disabling the gate
($\NoVerPart\%$) or context management ($\NoSusPart\%$) each produces a smaller drop.

\vspace{-.5em}
\paragraph{Delegation depth (Table~\ref{tab:ablation}b).}
We hold state-grounding fixed and vary depth: i)~flat delegation (default \ours),
ii)~worker recursion (depth $\leq$2), iii)~nested self-recursion with its own \gate. Flat
delegation leads on mean-partial ($\HarnessPart$ \textit{vs.}\ $\Bpart$ / $\Cpart$). The recursive
branch fired on only $7$ of $108$ tasks and never nested, so the between-config differences
cannot be attributed to depth. This supports a design choice (keep \ours flat), not a
universal claim about hierarchy.

\begin{table}[t]
\caption{Ablation on components and delegation depth. (a)~Component sensitivity (binary\,/\,mean-partial, \%):
each row removes one mechanism from full \ours; act-on-state matters most. (b)~Main agent
depth by capability (mean partial, \%): flat delegation, worker recursion, nested self-recursion.}
\vspace{-.5em}
\label{tab:ablation}
\begin{center}\footnotesize
\begin{minipage}[t]{0.32\textwidth}
{\centering(a) Component ablation\par\vspace{2pt}}
\scriptsize
\setlength{\tabcolsep}{3pt}
\begin{tabularx}{\linewidth}{@{}>{\raggedright\arraybackslash}Xcc@{}}
\toprule
{Config}\llap{\rotatebox{50}{{\phantom{Stream}}}} & \rotatebox{50}{{Binary}} & \rotatebox{50}{{Part.}} \\
\midrule
\textbf{\ours}                      & \textbf{\HarnessBin} & \textbf{\HarnessPart} \\
\quad$-$verify (no \gate)           & \NoVerBin & \NoVerPart \\
\quad$-$sustain (plan off)          & \NoSusBin & \NoSusPart \\
\quad$-$act (GUI; cua blocked)      & \NoActBin & \NoActPart \\
\bottomrule
\end{tabularx}
\end{minipage}\hfill
\begin{minipage}[t]{0.64\textwidth}
{\centering(b) Recursion scaffolds (per capability, mean partial)\par\vspace{3pt}}
\scriptsize
\setlength{\tabcolsep}{2pt}
\begin{tabular}{@{}lcccccccccc@{}}
\toprule
\textbf{Config} & \rotatebox{50}{M-item} & \rotatebox{50}{Stream} & \rotatebox{50}{X-src} & \rotatebox{50}{Confl} & \rotatebox{50}{Vis-sp} & \rotatebox{50}{Implic} & \rotatebox{50}{Tutor} & \rotatebox{50}{M-edit} & \rotatebox{50}{Dyn} & \rotatebox{50}{H-loop} \\
\midrule
\textbf{flat (\ours)} & \textbf{67} & \textbf{67} & \textbf{65} & \textbf{65} & \textbf{61} & \textbf{60} & \textbf{55} & \textbf{54} & \textbf{54} & \textbf{44} \\
worker                & 58 & 62 & 58 & 55 & 55 & 55 & 53 & 52 & 54 & 39 \\
nested                & 61 & 67 & 59 & 56 & 59 & 57 & 54 & 49 & 58 & 35 \\
\bottomrule
\end{tabular}
\end{minipage}
\end{center}
\end{table}

\subsection{Additional designs}
\label{sec:exp-transfer} Table~\ref{tab:transfer} reports additional designs. A bash-only configuration (no GUI, subagents, or \gate) reaches $\CodeOnlyPart\%$ partial, below even the reference
($\BasePart\%$): code-action alone is not enough without the scaffold. On Claude~Sonnet~4.6, the
same harness lifts binary success from $\TransBbBaseBin\%$ to $\TransBbBin\%$, evidence
that state-grounding helps a weaker backbone. On OSWorld-Verified, a short-horizon
benchmark, \ours and the reference perform similarly ($\TransBmBin\%$ \textit{vs.}\ $\TransBmBaseBin\%$
binary; Table~\ref{tab:transfer}c), consistent with state-grounding's advantage being concentrated on long-horizon tasks.

\begin{table}[t]
\caption{Additional system comparisons (binary / mean-partial, \%).
\textbf{(a)}~Opus~4.8 diagnostics:
bash-only exposes only a shell. \textbf{(b)}~Sonnet backbone transfer against the reported
reference-harness aggregate. \textbf{(c)}~OSWorld-Verified, a short-horizon benchmark (same Opus~4.8 backbone).}
\label{tab:transfer}
\begin{center}\footnotesize
\begin{minipage}[t]{0.44\textwidth}\centering
\textbf{(a)} Opus~4.8 diagnostics.\\[4pt]
\setlength{\tabcolsep}{8pt}
\begin{tabular}{@{}lcc@{}}
\toprule
System & Bin & Part \\
\midrule
Bash-only        & \CodeOnlyBin & \CodeOnlyPart \\
\textbf{\ours}            & \textbf{\HarnessBin} & \textbf{\HarnessPart} \\
\bottomrule
\end{tabular}
\hspace{-2em}
\end{minipage}\hfill
\begin{minipage}[t]{0.28\textwidth}\centering
\textbf{(b)} Sonnet~4.6.\\[4pt]
\setlength{\tabcolsep}{8pt}
\begin{tabular}{@{}lcc@{}}
\toprule
System & Bin & Part \\
\midrule
Reference & \TransBbBaseBin & \TransBbBasePart \\
\textbf{\ours}     & \textbf{\TransBbBin} & \textbf{\TransBbPart} \\
\bottomrule
\end{tabular}
\end{minipage}\hfill
\begin{minipage}[t]{0.28\textwidth}\centering
\textbf{(c)} OSWorld-Verified.\\[4pt]
\setlength{\tabcolsep}{8pt}
\begin{tabular}{@{}lcc@{}}
\toprule
System & Bin & Part \\
\midrule
Reference & \TransBmBaseBin & \TransBmBasePart \\
\textbf{\ours}     & \textbf{\TransBmBin} & \textbf{\TransBmPart} \\
\bottomrule
\end{tabular}
\end{minipage}
\end{center}
\end{table}

\section{Discussion}
\label{sec:discussion}
\subsection{Why State-Grounding Helps: A Failure Analysis}
\label{sec:diagnostic}

\begin{figure}[t]
\centering
\includegraphics[width=\linewidth]{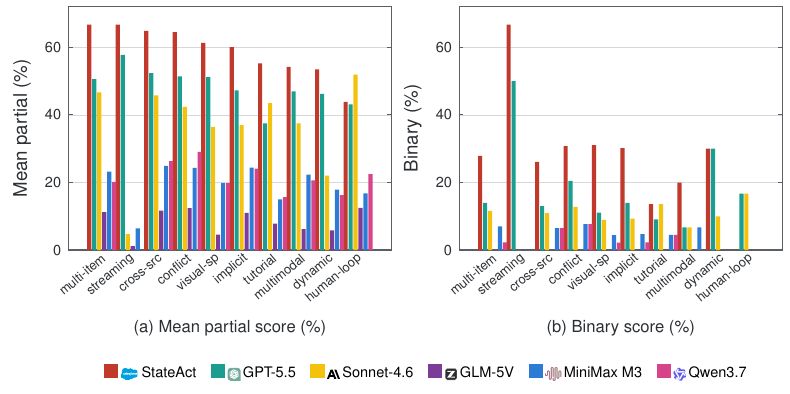}
\vspace{-2em}
\caption{Per-capability performance across six systems on OSWorld~2.0's ten
multi-label capability groups (\% within each group): (a) mean partial credit,
(b) binary success. \ours (red) uses Claude~Opus~4.8; public
points are from publicly released OSWorld~2.0 trajectories~\citep{osworld2}.
The best Opus-4.7 batch run and the same-backbone reference have no released
per-capability trajectories and are therefore omitted.}
\label{fig:failure}
\end{figure}

\begin{figure}[t]
\centering
\includegraphics[width=\linewidth]{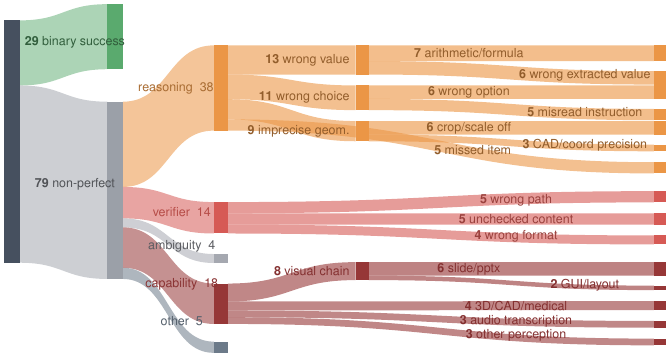}
\vspace{-1em}
\caption{Per-task dominant-cause audit of the $\NonPerfect$ non-perfect tasks. Reasoning
errors ($\FaReason$) dominate; $\sim\Ceiling\%$ hit modality bottlenecks or ambiguous
instructions. Categories are manual audit labels from trajectory inspection.}
\label{fig:errtax}
\end{figure}

Having established the gain (\S\ref{sec:experiments}), we ask \emph{where} the harness
pays off and \emph{what} remains. Figure~\ref{fig:failure} breaks scores down by capability
across six systems with available trajectories; Figure~\ref{fig:errtax} partitions the $\NonPerfect$ non-perfect
tasks by root cause. The pattern: \ours's largest margins fall on capabilities whose state
is machine-checkable (multi-item state, cross-source reasoning, conflict disambiguation),
while the weakest capabilities (human-in-the-loop, multimodal editing) need either an interaction turn the
harness never solicits or a flawless long visual chain. Averaging the ten capability rows equally gives $\sim$59 partial \textit{vs.}\ $\sim$27 binary (the /108 aggregate is $\HarnessPart$/$\HarnessBin$). This persistent partial-vs-binary gap reflects the verifier ceiling analyzed next.

\vspace{-.25em}
\paragraph{Two recoverable classes and a hard residual.}
The audit (Figure~\ref{fig:errtax}) groups most of the $\NonPerfect$ non-perfect tasks
into two recoverable classes: \emph{agent-reasoning} (a wrong value or misread
instruction, $\FaReason$ tasks, the dominant mode) and \emph{verifier-weak / wrong-path}
finishes that the \gate wrongly accepted ($\FaVerif$). These $\FaRecov$ tasks are
plausibly addressable (the errors are reasoning failures or verification misses). A further $\FaHard$ tasks ($\sim\Ceiling\%$ of the suite) either hit modality bottlenecks the backbone cannot cross (audio, video, real-time interaction) or carry ambiguous instructions with several valid readings. The remaining $\FaOther$ tasks were left undecomposed.
Acting on state removes most
perception failures: with code, ``read cell \texttt{B7}'' or ``list the calendar
store'' is exact. The \gate targets persistence failures: wrong-path
deliverables, unsaved edits, format mismatches. What neither addresses is reasoning, the dominant recoverable class ($\FaReason$ tasks).

\vspace{-.5em}
\paragraph{The verifier's ceiling.}
The \gate catches \emph{structural} defects (missing file, wrong path, format mismatch)
but cannot adjudicate \emph{value} correctness: re-deriving from the same source under the
same interpretation reproduces the agent's wrong answer. Of the $\NonPerfect$ non-perfect tasks, $\GateNonPerf$ reached the gate
(Table~\ref{tab:gate}); it correctly rejected only $\GateFail$ and wrongly passed
$\GatePass$, an error rate of $\GatePass/\GateNonPerf$ ($\approx$90\%).  This rate is high because the gate checks only structure: most of these passes fail on value or reasoning errors no structural check can catch, and only $\FaVerif$ are misses a structural check could have caught. This is not narration-conditioned agreement bias~\citep{agreementbias} (the gate is independent of the agent's narration) but a common-mode failure: shared interpretation of the source. The $\GateFail$ correct rejections are high-precision
structural catches that drive the bounded retries (Figure~\ref{fig:gate}). The ceiling is
intrinsic: without ground-truth labels, a prompt-level verifier bounds structural, not
value, correctness. This is the axis on which hard-coded per-task
verifiers~\citep{opencomputer2026} win, at the cost of generality.

\begin{table}[t]
\caption{Finish-gate operating characteristics. i)~Gate verdicts \textit{vs.}\ grader
outcomes (\BinarySuccessN\ binary successes, \NonPerfect\ non-perfect, 108 total; \GateNoFin\ non-perfect tasks never called
\texttt{finish} and thus never reached the gate). ii)~Operating rates of the \gate.}
\vspace{-.5em}
\label{tab:gate}
\begin{center}\footnotesize\setlength{\tabcolsep}{8pt}
\begin{minipage}[t]{0.5\textwidth}\centering
\textbf{(a)} Final verdict \textit{vs.}\ grader\\[4pt]
\scriptsize
\begin{tabular}{@{}lccc@{}}
\toprule
& Binary success & Non-perfect & Total \\
\midrule
PASS & \GateCorrectPass & \GatePass & 96 \\
FAIL & \GateCorrectFail & \GateFail & 9 \\
No finish called & 0 & \GateNoFin & \GateNoFin \\
\midrule
Total & \BinarySuccessN & \NonPerfect & 108 \\
\bottomrule
\end{tabular}
\end{minipage}\hfill
\begin{minipage}[t]{0.46\textwidth}\centering
\textbf{(b)} Operating rates\\[4pt]
\scriptsize
\begin{tabular}{@{}lcc@{}}
\toprule
& Fraction & Rate \\
\midrule
Correct rejections / all rejections & 8/9 & 88.9\% \\
Correct rejects / non-perfect reached & 8/76 & 10.5\% \\
Correct passes / binary successes & 28/29 & 96.6\% \\
Tasks retried & 28/105 & 26.7\% \\
Retries ending in binary success & 6/28 & 21.4\% \\
\bottomrule
\end{tabular}
\end{minipage}
\end{center}
\end{table}

\subsection{Is a strong GUI subagent necessary?}
\label{sec:exp-gui}
\ours quarantines GUI interaction to a subagent invoked on only $\Rgui\%$ of main-agent steps ($\GuiTasks$ of $108$ tasks), so the main agent rarely relies on screen-based control. Must that subagent be a frontier GUI model? We swap Claude's computer-use model for \pa{SFR-CUA}, our compact $31$B in-house computer-use model, holding Claude~Opus~4.8 as the main agent in both, and evaluate across five benchmarks (Table~\ref{tab:cross-benchmark}; the parenthetical names the GUI subagent, and Claude~Opus~4.8 is the reference).

On four of the five benchmarks the substitution barely moves the end-to-end mean-partial score: OSWorld-Verified ($81.1$ \textit{vs.}\ $\TransBmPart$), WindowsAgentArena ($51.2$ \textit{vs.}\ $50.6$), AndroidWorld ($84.1$ \textit{vs.}\ $81.9$), and MobileWorld ($68.4$ \textit{vs.}\ $70.1$). This holds even though SFR-CUA on its own is far weaker than Claude~Opus~4.8, scoring $66.9$ \textit{vs.}\ $80.9$ on OSWorld-Verified and $7.6$ \textit{vs.}\ $54.8$ on OSWorld~2.0. Once the main agent carries the task on state, a compact specialist suffices for the rare visual fallback. The exception is OSWorld~2.0, our longest-horizon suite: there \ours(SFR-CUA) reaches only $43.2\%$ partial and $18.5\%$ binary, below \ours(Claude~Opus~4.8) at $\HarnessPart\%$/$\HarnessBin\%$ and the base model at $\BasePart\%$/$\BaseBin\%$, because its harder visual subgoals expose the weaker subagent. A frontier GUI model is thus unnecessary on the shorter-horizon and mobile benchmarks but still helps on the hardest long-horizon desktop tasks.

\begin{table}[t]
\caption{Performance across five benchmarks (mean-partial score, \%):
OSWorld-Verified~\citep{osworld}, OSWorld~2.0~\citep{osworld2}, WindowsAgentArena~\citep{windowsarena}, AndroidWorld~\citep{androidworld}, and
MobileWorld~\citep{mobileworld}.}
\label{tab:cross-benchmark}
\vspace{-.5em}
\begin{center}\footnotesize\setlength{\tabcolsep}{2pt}
\small
\begin{tabular*}{\linewidth}{@{\extracolsep{\fill}}lccccc@{}}
\toprule
\textbf{Model} & \shortstack{OSWorld\\Verified} & \shortstack{OSWorld\\2.0} &
\shortstack{WindowsAgent\\Arena} & \shortstack{Android\\World} & \shortstack{Mobile\\World} \\
\midrule
Claude~Opus~4.8 & 80.9 & 54.8 & 41.6 & 69.0 & 51.3\\
SFR-CUA               & 66.9 & 7.6 & 40.9 & 68.1 & 48.7 \\
\ours\,(Claude~Opus~4.8)      & 81.9 & 61.6 & 50.6 & 81.9 &70.1  \\
\ours\,(SFR-CUA)      & 81.1 & 43.2 & 51.2 & 84.1 & 68.4 \\
\bottomrule
\end{tabular*}
\end{center}
\end{table}

\section{Conclusion}
\label{sec:conclusion}
We presented \ours, a harness that makes program state as the  primary interface for the main agent  while retaining a dedicated GUI subagent for visual interaction. On a standard long-horizon GUI benchmark, Claude~Opus~4.8 rises from $\BaseBin\%$ to $\HarnessBin\%$ binary success ($\BasePart\%$ to $\HarnessPart\%$ partial) at $\sim\CostRatio\times$ lower cost, without any change to the model itself. The ablation and diagnostic analyses localize both the gain and its limit: the gain comes from \emph{what} the agent observes (state rather than screenshots), not from added depth; the limit is value correctness, which a self-verifier without ground-truth labels cannot close. State-grounding moves the accuracy wall from perception to reasoning; it does not remove it. For long-horizon computer use, the bottleneck is now what the agent \emph{thinks}, not what it \emph{sees}.

\bibliography{iclr2025_conference}
\bibliographystyle{iclr2025_conference}

\end{document}